\UseRawInputEncoding
\documentclass[epj]{svjour}
\usepackage{graphics}
\begin{document}
\title{Measurement of the $E_{\rm p}$\,=\,416.9\,keV resonance strength in the $^{29}$Si(p,$\gamma$)$^{30}$P reaction}
\author{Zs. M\'atyus\inst{1,2} \and L. Csedreki\inst{1} \and Zs. F\"ul\"op\inst{1} \and Z. Hal\'asz\inst{1} \and G.G. Kiss\inst{1} \and T. Sz\"ucs\inst{1} \and \'A. T\'oth\inst{1,2} \and Gy. Gy\"urky\inst{1}%
\thanks{Corresponding author, gyurky@atomki.hu}%
}                     % Do not remove
\institute{HUN-REN Institute for Nuclear Research (ATOMKI), H-4001 Debrecen, Hungary \and University of Debrecen, Doctoral School of Physics, Egyetem t\'er 1., 4032 Debrecen, Hungary}
\date{Received: date / Revised version: date}
\abstract{Silicon isotopic ratios measured in meteoritic presolar grains can provide useful information about the nucleosynthesis origin of these isotopes if the rates of nuclear reactions responsible for their production are known. One of the key reactions determining the Si isotopic abundances is $^{29}$Si(p,$\gamma$)$^{30}$P. Its reaction rate is not known with sufficient precision due in part to some ambiguous resonance strength values. In the present work, the strength of the $E_{\rm p}$\,=\,416.9\,keV resonance has been measured with high precision using the activation technique. The new strength of $\omega\gamma$\,=\,219\,$\pm$\,16\,meV can be used in updated reaction rate estimations and astrophysical models.
 % end of PACS codes
} %end of abstract
\maketitle
\section{\label{sec:intro} Introduction}

Classical novae, the thermonuclear runaway events on the surface of accreting white dwarfs \cite{Jose2016}, are thought to be prolific sources of interstellar dust. This dust, produced before the formation of the Solar System, can be found in presolar grains buried in primitive meteorites. Isotopic ratios measured in such presolar grains can provide insight into the nucleosynthesis processes in a nova explosion \cite{Jose2007}, if such grains can be unambiguously associated with nova events. Up to now, however, no grain could be clearly proven to have classical nova origin as supernovae are often suggested as an alternative source \cite{Downen2022b}.

One reason of the missing evidence is the insufficient knowledge of thermonuclear reaction rates of those nuclear reactions which play a role in the production or destruction of a given isotope. In the case of silicon isotopes, for example, the $^{29}$Si(p,$\gamma$)$^{30}$P and $^{30}$P(p,$\gamma$)$^{31}$S reactions determine mainly the $^{28}$Si:$^{29}$Si and $^{28}$Si:$^{30}$Si ratios, respectively \cite{Downen2022b}. A sensitivity study of C. Iliadis \textit{et al.} \cite{Iliadis2002} indicate that the rate uncertainties of the $^{29}$Si(p,$\gamma$)$^{30}$P and $^{30}$P(p,$\gamma$)$^{31}$S reactions strongly contribute to the uncertainty of Si isotopic ratio predictions in classical novae. The topic of the present work is the study of one of these reactions, $^{29}$Si(p,$\gamma$)$^{30}$P, therefore, the experimental information on this reaction is summarized below and the need for new measurements is emphasized.

Recently, L. N.  Downen \textit{et al.} measured the strength of some low-energy resonances in $^{29}$Si(p,$\gamma$)$^{30}$P and provided updated thermonuclear reaction rate values \cite{Downen2022}. Depending on the temperature, their reaction rates have an uncertainty of up to an order of magnitude. At the lowest temperatures (below about 0.025\,GK) the rate is dominated by the direct capture reaction, for which no experimental data exists in the literature. Above this temperature, up to 2\,GK, low energy resonances, studied in \cite{Downen2022} give the most important contribution to the rate. In their work, L. N.  Downen \textit{et al.} did not measure the strength of the relatively strong $E_{\rm p}$\,=\,416.9\,keV resonance\footnote{In this paper the resonance energies are given in the laboratory system as proton energies, calculated from the known level energies, atomic masses and reaction Q-value \cite{Downen2022}.}, but they accepted a value from the literature and carried out a relative measurement, normalizing the measured strengths of other resonances to this resonance. Therefore, the uncertainty of the 416.9\,keV resonance strength contributes directly to the uncertainties of the other measured resonance strengths and hence of the reaction rate.

The strength of the 416.9\,keV resonance was measured several times but all these measurements date back to many decades ago. Table \ref{tab:literature} summarizes the experimental data. As it can be seen, differences of almost a factor of five are found, which renders the strength value rather unreliable. L. N.  Downen \textit{et al.} \cite{Downen2022} adopts the value of $\omega\gamma$\,=\,220\,$\pm$\,25\,meV based on the list given in \cite{Sargood1982}. This value is smaller by about 20\,\% than the value in the latest compilation \cite{Endt1998} and much higher than the results of some early measurements.

\begin{table*}
\caption{\label{tab:literature} Measured and compiled values for the $E_{\rm p}$\,=\,416.9\,keV resonance strength}
\begin{tabular}{llcl}
\hline
Year & Reference & Resonance  & Remark \\
 				&						& strength [meV] 					&  \\
\hline

1956 & C. Broude \textit{et al.} \cite{Broude1956} & 57.5 & cited by \cite{vanderLeun1958} without error \\
1958 & C. van der Leun \textit{et al.}  \cite{vanderLeun1958} & 58.5\,$\pm$\,15 \\
1966 & G. A. P. Engelbertink \textit{et al.}  \cite{Engelbertink1966} & 175\,$\pm$\,25 \\
1979 & M. Riihonen \textit{et al.}  \cite{Riihonen1979} & 260\,$\pm$\,25 \\
\hline
1982 & D. G. Sargood \cite{Sargood1982} & 220\,$\pm$\,25 & rescaling \cite{Engelbertink1966} based on \cite{Paine1979}, \\
 & & & adopted by \cite{Downen2022} \\
1998 & P. M. Endt  \cite{Endt1998} & 260\,$\pm$\,25 & most recent compilation, \\
& & & adopts \cite{Riihonen1979} \\
\hline
\end{tabular}
\end{table*}

Motivated by these ambiguities and by the fact that the strength of the 416.9\,keV resonance has not been measured since more than four decades, the purpose of the present work was to carry out a new measurement for this resonance strength. 

In section \ref{sec:experimental} the details of the experiment will be provided followed by the discussion of the data analysis in section \ref{sec:analysis}. The results will be summarized in section \ref{sec:summary}.

\section{\label{sec:experimental} Experimental procedure}

In the present experiment the activation technique was applied. In this method, the resonance strength measurement is based on the detection of the decay of the \linebreak $^{29}$Si(p,$\gamma$)$^{30}$P reaction product which is radioactive and decays to $^{30}$Si with a half-life of 2.498\,$\pm$\,0.004 min \cite{Basunia2010}. The positron emission of $^{30}$P is not followed by any $\gamma$ radiation, but the decay can be observed by the detection of the 511\,keV positron annihilation radiation. This method was applied in the present work, which has the advantage that it provides the astrophysically important total resonance strength, independently from the decay pattern of the populated level in $^{30}$P. 

\subsection{Target properties} 
To determine the strength of the $E_{\rm p}$\,=\,416.9\,keV resonance, 3\,mm thick SiO$_2$ targets, purchased from the Kurt J Lesker company (part no. EJTSIO2451A2), were used having high (99.995\,\%) chemical purity. Our first attempt to use thick elemental Si targets failed because no stable and reproducible reaction yield could be achieved. 
%SiO$_2$ has been chosen as target material as the Si:O ratio of 1:2 is stable and well know.
Since SiO$_2$ is not a good electrical conductor, thin Al layers were evaporated on the targets to drain the electric charge deposited by the proton beam. Based on the evaporation geometry, the thickness of the Al layer was such that the proton energy loss in it was about 2\,keV at $E_{\rm p}$\,=\,416.9\,keV. Altogether four targets of the same type were used. 

\subsection{Irradiation} 
The proton beam was provided by the Tandetron accelerator of ATOMKI \cite{Rajta2018,Biri2021}. The beam current was typically about 2.5-3\,$\mu$A. At the entrance of the target chamber an electrode biased at -300\,V was placed to suppress the secondary electrons emitted from the beam defining collimator or from the target. The target chamber behind this electrode served as a Faraday cup to measure the electric charge carried by the beam to the target. From the charge, measured by an ORTEC 439 current integrator, the number of protons impinging on the target can be derived. The current integrator was calibrated against a Keithley 6748 picoammeter in the actual experimental conditions. The current was integrated for 5 seconds intervals. This allowed to monitor and register possible beam intensity variations during the irradiation. The beam intensity was relatively stable during the irradiation periods (typical fluctuations not more that a few percent) but even these fluctuations were taken into account in the data analysis based on the recorded integrator counts. The applied beam energies were in the range between $E_{\rm p}$\,=\,394-433\,keV, below and above the resonance energy (see Sec.\,\ref{sec:analysis}).  

Owing to the short half-life of the reaction product, the cyclic activation method was used, meaning the irradiation and decay counting phases being repeated many times. The duration of the irradiation in one cycle was 5\,min, followed by a 15\,min beam-off period for decay counting. In one experimental run (i.e. at a given proton energy, on a given target, without interrupting the series of cycles), many cycles up to 54 were completed to improve the statistical accuracy of the collected data.

Altogether eight multi-cycle runs were carried out and analysed, six run above, and two below the resonance energy. Table \ref{tab:yields} provides some details of these runs (in time order) including the target number, the proton beam energy, the number of cycles and the obtained yields (see Sec.\,\ref{sec:analysis}).  

\subsection{Detection of the $^{30}$P decay} 
During both the irradiation time and the decay time, the 511\,keV $\gamma$ radiation was detected by a HPGe detector. Since the half-life of $^{30}$P is relatively short, the decay was measured without removing the targets from the irradiation chamber. The HPGe detector was placed behind the chamber with its front face about 20\,mm far from the target. The chamber and detector arrangement can be seen in Fig. 2 of \cite{Gyurky2023}. Similarly to the charge integrator counts, the 511\,keV events were also recorded in 5\,sec time intervals. Owing to the low count rate (always less than 100 count/sec), the dead time of the acquisition system was negligible.

The absolute efficiency measurement of the HPGe detector in the counting geometry was carried out with calibration sources and using the $^{12}$C(p,$\gamma$)$^{13}$N reaction, as described in detail in \cite{Gyurky2019}. First the efficiency was determined in a reference geometry using calibrated radioactive sources. Then a $^{13}$N source was produced with the $^{12}$C(p,$\gamma$)$^{13}$N reaction which decays by positron emission, similarly to $^{30}$P, but with a longer half-life of 9.965\,min. This longer half-life allows the measurement of the $^{13}$N decay first in the counting geometry (when the target is in the irradiation chamber) and then in the reference geometry (after removing the target from the chamber). From the comparison of the two measurements the absolute efficiency in the counting geometry can be determined. With this procedure the absolute detection efficiency was found to be $\eta$\,=\,(1.83\,$\pm$\,0.05)\%.

\begin{table*}
\caption{\label{tab:yields} Some parameters of the multi-cycle runs. The quoted yield uncertainties are statistical only. See text for further details.}
\begin{tabular}{ccccrcl}
\hline
Run no. (type) & Target & $E_{\rm p}$ [keV]  & No. of cycles & \multicolumn{3}{c}{Yield $\times  10^{-15}$} \\
\hline

\# 1 (on res.)	&	SiO$_2$ - 3	&	433.2	&	24	&	4777	&	$\pm$	&	189	\\
\# 2 (on res.)	&	SiO$_2$ - 3	&	433.2	&	54	&	4340	&	$\pm$	&	121	\\
\# 3 (on res.)	&	SiO$_2$ - 1	&	433.2	&	20	&	4068	&	$\pm$	&	193	\\
\# 4 (on res.)	&	SiO$_2$ - 2	&	433.2	&	48	&	4502	&	$\pm$	&	147	\\
\# 5 (off res.)	&	SiO$_2$ - 3	&	403.8	&	46	&	1018	&	$\pm$	&	132	\\
\# 6 (on res.)	&	SiO$_2$ - 4	&	427.3	&	15	&	4069	&	$\pm$	&	306	\\
\# 7 (on res.)	&	SiO$_2$ - 4	&	433.2	&	11	&	4703	&	$\pm$	&	351	\\
\# 8 (off res.)	&	SiO$_2$ - 4	&	394.0	&	32	&	873	&	$\pm$	&	154	\\
\hline													
\multicolumn{4}{l}{$Y_{\rm tot.}$}							&	4406	&	$\pm$	&	109	\\
\multicolumn{4}{l}{$Y_{\rm off}$}							&	957	&	$\pm$	&	100	\\
\multicolumn{4}{l}{$Y_{\rm res.}$}							&	3450	&	$\pm$	&	148	\\
\hline
\end{tabular}
\end{table*}

\section{\label{sec:analysis} Data analysis}

\begin{figure}[h!]
    \resizebox{1.45\columnwidth}{!}{\includegraphics{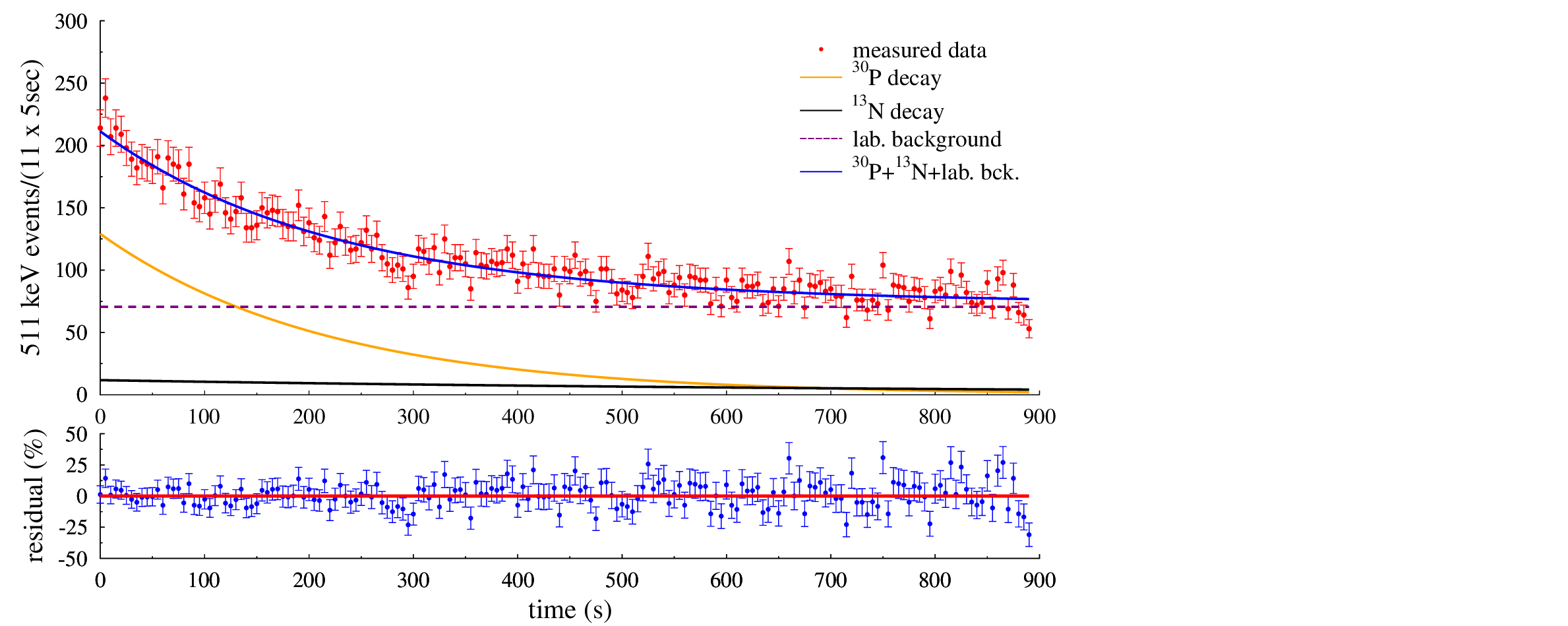}}
    \caption{Number of detected 511\,keV $\gamma$ rays as a function of time and the fit of the data including three components as described in the text. The shown decay curve is the sum of 11 cycles of the run \# 7 with proton energy of 433.2\,keV.}
    \label{fig:decay}
\end{figure}

The decay periods of the many-cycle activation runs were summed and the analysis of the decay curves was used to calculate the resonance strength. Such a decay curve can be seen in Fig.\,\ref{fig:decay} corresponding to a run of 11 cycles on target SiO$_2$ - 4 irradiated with 433.2\,keV protons (run \#7 in Table\,\ref{tab:yields}). The decay of $^{30}$P is not the only possible source of the 511\,keV $\gamma$ radiation. The laboratory background must be taken into account as well as the decay of $^{13}$N. This isotope is created by the $^{12}$C(p,$\gamma$)$^{13}$N reaction on the carbon contaminants of the target or collimator. The longer half-life of $^{13}$N (9.965\,min compared to the 2.498\,min half-life of $^{30}$P) allows the separation of the two sources by fitting the decay curve as the sum of two exponentials and a constant background (three free parameters). Such a fit can be seen in the figure. 
From the fit the number of the produced $^{30}$P nuclei and thus the number of $^{29}$Si(p,$\gamma$)$^{30}$P reactions could be derived. Using the recorded current integrator counts, the reaction yield $Y=N_{\rm r}/N_{\rm p}$, defined as the ratio of the number of reactions ($N_{\rm r}$) and the number of projectiles bombarding the target ($N_{\rm p}$), could be determined, which is related to the strength of the resonance. The standard formulae that connect the measured number of counts to the reaction yield in an activation experiment can be found e.g. in refs. \cite{gyurky2019b,diLeva2014}.

Since the 3\,mm thick target completely stops the proton beam, the thick target assumption can be used to obtain the resonance strength $\omega$$\gamma$, using the following standard formula \cite{Iliadis2007}: 
\\
\begin{equation} \label{eq:res_str}
    \omega\gamma\,=\frac{2\epsilon_{\rm eff}Y_{\rm res.}}{\lambda^{2}},
\end{equation}
where $\lambda$ is the de-Broglie wavelength at the resonance energy in the center-of-mass system, $\epsilon_{\rm eff}$ is the effective stopping power of the SiO$_2$ target for protons at the resonance energy and $Y_{\rm res.}$ is the contribution of the studied resonance to the measured yield. In the case of an infinitely thick target, the measured yield $Y_{\rm tot.}$ above the resonance is the sum of the resonance yield $Y_{\rm res.}$ and the reaction yield $Y_{\rm off}$ below the resonance energy (either direct capture or lower energy resonances).

In order to calculate $Y_{\rm res.}$, six measurements were carried out above the resonance energy (at $E_{\rm p}$\,=\,427 or 433 keV, to obtain $Y_{\rm tot.}$) and two measurements below the resonance (at $E_{\rm p}$\,=\,396 and 404\,keV for $Y_{\rm off}$), as listed in Table\,\ref{tab:yields}. The weighted average of the six measurement above the resonance was calculated to obtain $Y_{\rm tot.}$. The reduced chi square value of the averaging is 2.218, i.e. higher than one. Therefore, the uncertainty of the average is multipiled by the square root of $\chi^2_{\rm red.}$. The obtained yield is thus $Y_{\rm tot.}$\,= (4406\,$\pm$\,109)$\times 10^{-15}$. The two yields below the resonance are statistically fully consistent and their weighted average is calculated to give $Y_{\rm off}$\,=(957\,$\pm$\,100)$\times 10^{-15}$. The difference of these two values gives $Y_{\rm tot.}$\,-$Y_{\rm off}$\,=\,$Y_{\rm res.}$\,= \linebreak (3450\,$\pm$\,148)$\times 10^{-15}$. Here the uncertainties are statistical only, determined from the fit of the decay curves as discussed above. 

%The weighted averages of these measured yields are: $Y_{\rm tot.}$\,= \linebreak (4406\,$\pm$\,109)$\times 10^{-15}$ and $Y_{\rm off}$\,=(957\,$\pm$\,100)$\times 10^{-15}$. The difference of these values gives $Y_{\rm res.}$\,=(3450\,$\pm$\,148)$\times 10^{-15}$. Here the uncertainties are statistical only, determined from the fit of the decay curves as discussed above. 

The effective stopping power entering equation\,\ref{eq:res_str} is calculated from the stopping powers of Si and O obtained from the SRIM code \cite{SRIM} using the following formula, where $N_{\rm x}$ stands for the density of the component x in the target \cite{Iliadis2007}:

\begin{equation} \label{eq:stopping}
\epsilon_{\rm eff}=\epsilon_{\rm Si}\Biggl(1+\frac{N_{^{28}{\rm Si}+^{30}{\rm Si}}}{N_{^{29}{\rm Si}}}\Biggr) + \epsilon_{\rm O}\frac{N_{\rm O}}{N_{^{29}{\rm Si}}}. 
\end{equation}

The values for the Si and O stopping powers at 416.9\,keV resonance energy, taken from SRIM, are: $\epsilon_{\rm Si}$=13.1$\pm$\,1.1 and $\epsilon_{\rm O}$=9.10$\pm$\,0.21, in units of eV/10$^{15}$\,atoms/cm$^2$. Here the uncertainties are adopted from the SRIM experimental stopping power plots \cite{SRIMerror}. 3\% relative uncertainty is assigned to the the 1/3 stoichiometric ratio of Si in SiO$_2$. The isotopic abundance of $^{29}$Si in natural Silicon is known with high accuracy: (4.685\,$\pm$\,0.008)\%. 

Table \ref{tab:uncer} summarizes the sources of uncertainties on the determined resonance strength. Other factors, like the de-Broglie wavelength or the isotopic abundance of $^{29}$Si have uncertainties well below 1\% and thus neglected. The values for the measured yields are obtained from the decay curve fit as discussed above. The total uncertainty is calculated as the quadratic sum of the factors listed in the last column. 

\begin{table}[h!]
\centering
\caption{\label{tab:uncer} Uncertainty sources of the measured $E_{\rm p}$\,=\,416.9\,keV resonance strength}
\begin{tabular}{llcl}
\hline
Source &  \multicolumn{2}{c}{relative uncertainty} \\
\hline
$Y_{\rm tot.}$ (statistical only) & 2.5\,\% &    \\
$Y_{\rm off}$ (statistical only) & 10.5\,\% &   \\
%\cline{1-2}
$Y_{\rm res.}=Y_{\rm tot.}-Y_{\rm off}$ (statistical only) & &    4.3\%\\
\cline{1-2}
Si stopping power & 8.0\,\% &   \\
O stopping power & 2.3\,\% &   \\
Si:O ratio & 3.0\,\% &   \\
%\cline{1-2}
Effective stopping power & & 4.5\%\\
\cline{1-2}
HPGe detector efficiency & & 3.0\%\\
\cline{1-2}
Current integration & & 3.0\%\\
\hline
Total &&7.5\%\\
\hline
\end{tabular}
\end{table}

\section{\label{sec:summary} Results and conclusions}
With the described experimental technique and data analysis, the following strength is obtained for the \linebreak $E_{\rm p}$\,=\,416.9\,keV resonance in the $^{29}$Si(p,$\gamma$)$^{30}$P reaction: \linebreak $\omega\gamma$\,=\,219$\pm$16\,meV. 
This result is in perfect agreement with the value quoted by D. G. Sargood \cite{Sargood1982} and used by L. N.  Downen \textit{et al.} \cite{Downen2022}, but with smaller uncertainty. 

This new precise resonance strength can be used to recalculate the strengths of the lower energy resonances measured by L. N.  Downen \textit{et al.} \cite{Downen2022} as those strengths were normalized to the $E_{\rm p}$\,=\,416.9\,keV resonance studied in the present work. The result of this recalculation can be found in Table \ref{tab:results}.

\begin{table}[h!]
\caption{\label{tab:results} Strengths of the low energy resonances in $^{29}$Si(p,$\gamma$)$^{30}$P. The strength of the $E_{\rm p}$\,=\,416.9\,keV resonance is determined in the present work and the lower energy resonance strengths are rescaled from the work of L. N.  Downen \textit{et al.}}
\begin{tabular}{lcc}
\hline
$E_{\rm p}$ & \multicolumn{2}{c}{$\omega\gamma$ [meV]} \\
\cline{2-3}
\hspace{0mm}[keV] & L. N.  Downen \textit{et al.} \cite{Downen2022} & present work \\
\hline
313.9 & 0.088\,$\pm$\,0.015 &  0.088\,$\pm$\,0.013$^1$ \\
325.1 & 20.7\,$\pm$\,2.7 &  20.6\,$\pm$\,2.0$^1$ \\
416.9 & 220\,$\pm$\,25$^2$ &  219\,$\pm$\,16  \\
\hline
\end{tabular}\\
$^1$ rescaled from \cite{Downen2022}\\
$^2$ adopted from \cite{Sargood1982}
\end{table}

As the resonance strength values do not change significantly, the astrophysical reaction rates determined in \cite{Downen2022} are also unchanged and the astrophysical conclusions remain the same. At low temperatures, the direct capture component of the $^{29}$Si(p,$\gamma$)$^{30}$P reaction - for which no experimental data is available in the literature - gives the most important contribution to the reaction rate. Since the activation method applied in the present work proved to be a suitable technique for studying the $^{29}$Si(p,$\gamma$)$^{30}$P reaction, as a continuation of the present experiment the direct capture cross section is also measured with activation. The results of this ongoing experiment will be presented in a forthcoming publication.

\section*{Acknowledgments}
This work was supported by NKFIH grant K134197, by the \'UNKP-23-3 New National Excellence Programs of the Ministry for Culture and Innovation from the source of the National Research, Development and Innovation fund and by the Hungarian
Government, Economic Development and Innovation Operational Programme (GINOP-2.3.315-2016-00005) grant, co-funded by the EU.

% BibTeX users please use
%\bibliographystyle{epj}
%\bibliography{Matyus_29Si_bib}
%\begin{thebibliography}{}


\end{document}